\documentclass[prd,groupedaddress, showpacs, showkeys, twocolumn, amsmath,amssymb,nofootinbib]{revtex4}

\usepackage{graphicx}% Include figure files
\usepackage{dcolumn}% Align table columns on decimal point
\usepackage{bm}% bold math
\usepackage{amsmath,graphics,color}
\usepackage{color,psfrag}
\usepackage{mathrsfs}

\newcommand{\nn}{\nonumber}
\newcommand{\be}{\begin{equation}}
\newcommand{\ee}{\end{equation}}
\newcommand{\bea}{\begin{eqnarray}}
\newcommand{\eea}{\end{eqnarray}}

\newcommand{\vS}{\widetilde{\mbox{\boldmath$\epsilon$}}}

\newcommand{\norm}{| \mspace{-2mu} |}

\begin{document}

\def\la{\langle}
\def\ra{\rangle}
\def\a{\alpha}
\def\b{\beta}
\def\g{\gamma}
\def\d{\delta}
\def\e{\epsilon}
\def\phi{\varphi}
\def\k{\kappa}
\def\l{\lambda}
\def\m{\mu}
\def\n{\nu}
\def\o{\omega}
\def\p{\pi}
\def\r{\rho}
\def\s{\sigma}
\def\t{\tau}
\def\del{\partial}
\def\nab{\nabla}

\title{Non-equilibrium Thermodynamics of Spacetime:\\ the Role of Gravitational Dissipation}

%-------------------------------------------------------------------------
\author{G. Chirco}
\email{chirco@sissa.it}
%-------------------------------------------------------------------------
\author{S. Liberati}
\email{liberati@sissa.it}
%-------------------------------------------------------------------------
% \email{chirco@sissa.it}
 \affiliation{SISSA,  Via Beirut 2, 34151 Trieste, Italy and INFN sezione di Trieste}
%\affiliation{${}^2$Universit\'a di Udine, Via delle Scienze 208, 33100 Udine, Italy}
%-------------------------------------------------------------------------
\begin{abstract}
In~\cite{Jacobson:1995ab} it was shown that the Einstein equation can be derived as a local constitutive equation for an equilibrium spacetime thermodynamics.
More recently, in the attempt to extend the same approach to the case of $f(R)$ theories of gravity, it was found that a non-equilibrium setting is indeed required in order to fully describe both this theory as well as classical GR~\cite {Eling:2006}. Here, elaborating on this point,  we show that the dissipative character leading to a non-equilibrium spacetime thermodynamics is actually related --- both in GR as well as in $f(R)$ gravity --- to non-local heat fluxes associated with the purely gravitational/internal degrees of freedom of the theory. In particular, in the case of GR we show that the internal entropy production term is identical to the so called tidal heating term of Hartle-Hawking. Similarly, for the case of $f(R)$ gravity, we show that dissipative effects can be associated with the generalization of this term plus a scalar contribution whose presence is clearly justified within the scalar-tensor representation of the theory. Finally, we show that the allowed gravitational degrees of freedom can be fixed by the kinematics of the local spacetime causal structure, through the specific Equivalence Principle formulation. In this sense, the thermodynamical description seems to go beyond Einstein's theory as an intrinsic property of gravitation.
\\
\begin{center}
{\em This paper is dedicated to the memory of Dennis William Sciama} 
\end{center}
\end{abstract}
%-------------------------------------------------------------------------
\pacs{04.70.Dy; 04.20.Cv; 04.62.+v}
%-------------------------------------------------------------------------
\keywords{emergent gravity, black holes, thermodynamics, Einstein field equations, scalar-tensor gravity}
%-------------------------------------------------------------------------
\maketitle

\section{Introduction}
Over a decade ago, the connection between gravity, thermodynamics and quantum field theory, developed in 70's for black holes physics \cite{Bekenstein:1973ur,Bardeen:1973gs, Haw}, was strongly tightened up by Jacobson~\cite{Jacobson:1995ab}, who was able to derive the Einstein equations as equilibrium constitutive equations for spacetime, starting from the thermodynamical properties of local causal horizons and the thermal nature of the Minkowski vacuum. 

Since then, this result has given great support to the idea that the well established black holes thermodynamics should be in fact extendable to some more general spacetime thermodynamics, where the fundamental ingredients are a mix between causal horizons stationarity, quantum fields thermal behaviour and peculiar holographic properties of gravity \cite{Pad1,Para, Sotiriou:2006gp,Pad2, tof, sus}.
 
However, more recently, it was realized that the thermodynamical derivation of the Einstein equations needs a generalization to a non-equilibrium thermodynamical setting~\cite{Eling:2006, Eling:2008}. This was firstly noticed in the attempt to reproduce the derivation \`a la Jacobson for a $f(R)$ gravity theory. In that case, making the horizon entropy proportional to a function of the Ricci scalar led to a break down of the local thermodynamical equilibrium. 

In order to recover the $f(R)$ gravity field equations from the thermodynamical prescription, it was then necessary to modify the equilibrium entropy balance of the system by considering some extra entropy production term. More surprisingly, the same problem was in fact pointed out, in the same works, even for GR, the main issue being again substantially related to the definition of entropy for the system. Nevertheless, a clear physical interpretation of the extra entropy production terms which come into play for the non-equilibrium thermodynamics derivation is still missing.

In this work, following~\cite{Eling:2006}, we adopt the non-equilibrium description as the general proper setting for the above approach. In such non-equilibrium setting, the Einstein equations (or their generalization for higher order gravity) can be still derived from a local thermodynamical condition, as a constitutive equation, but only as far as one separates the reversible and irreversible thermodynamical contributions (with the reversible sector being associated with the Einstein field equations). In this context, the general form for the irreversible/viscous contributions and the constitutive equations for their coefficients is identified by exploiting a fluid dynamics description of the local causal horizon kinematics. We then argue that such terms, already interpreted in~\cite{Eling:2006} as dissipative effects, should be identified as heat terms associated with the purely gravitational/internal degrees of freedom of the theory. In particular, for the case of GR we find that the internal entropy production term is exactly the well known tidal heating term associated with dissipation of black hole horizons perturbations via gravitational fluxes~\cite{TP, HH, Chandra, Poi:2004, Poi:2005}. Furthermore, by applying the same approach to the $f(R)$ theories of gravity, we actually find a generalization of the tidal heating effect of GR together with an extra purely scalar dissipative contribution (as expected in this class of theories given their possible representation as scalar-tensor ones~\cite{Sotiriou:2008rp}).

We start, in section \ref{sec:framework}, by reviewing and analyzing the operative framework for the spacetime thermodynamics approach. In particular, we further elaborate on the role of the equivalence principle (EP) in selecting gravitational theories in such a framework.  We then proceed in section \ref{ees} to the derivation of the Einstein equation of state as in~\cite{Jacobson:1995ab, Eling:2006, Eling:2008}. In section \ref{ne-th}, we describe the analogy between a general fluid and the horizon congruence kinematics. This allows us to interpret (in a classical irreversible thermodynamics setting) the non-equilibrium features of the congruence as the analogues of the viscous dissipative terms for the fluid. In section \ref{neos} and \ref{sec:fR}, we re-derive the equations of motion for GR and $f(R)$ gravity respectively, by starting from a general non-equilibrium entropy balance and looking at its reversible sector. We show that the irreversible contributions can in both cases be associated with the purely gravitational dissipative effects of the theory.
In the Discussion we shall finally speculate on the possible lessons one can gather from our results and their hints towards an identification of the microscopic degrees of freedom of gravity.

\section{Spacetime thermodynamics: the Framework}\label{sec:framework}

The idea of a spacetime thermodynamics, as firstly presented in \cite{Jacobson:1995ab}, is built on the relation between the thermal character of quantum fields vacuum, as perceived by an accelerated Rindler observer, and the stationarity properties of the respective Killing horizon. By making use of the equivalence principle, the notion of Rindler frame can be used at a local level as an experimental setting for studying the local spacetime dynamics as the way the variation of the spacetime geometry follows from the energy variation of the matter fields. Here we review and extend the framework of the previous investigations~\cite{Jacobson:1995ab, Eling:2006, Eling:2008} stressing its crucial points and its implicit assumptions.

\subsection{The Local Rindler Wedge} \label{da}

The first step in the construction of a spacetime thermal system is the introduction of a local notion of horizon. 
In analogy with the global definition of a horizon as the boundary of the past of future null infinity, one can generally consider a local horizon at $p$, in a generic spacetime $(\mathcal{M}, g_{ab})$, as one side of the boundary of the past of a spacelike 2-surface patch $B$ including $p$. Thereby, near $p$, the local horizon will be constituted by the congruence of null geodesics orthogonal to $B$, characterized by the past pointing tangent null vector $\ell^a$ (see Fig.~1).

With respect to the point $p$, one can then invoke Local Lorentz Invariance of spacetime (assumed in both the Einstein and Strong formulations of the Equivalence Principle~\cite{Willl}) to introduce a local inertial frame (LIF).  This is always allowed, provided one restricts oneself to a region of size $\ell^2 \ll  {{\cal R}(p)}^{-1}$, where ${\cal R}(p)$ gives the value of the smaller scale associated with the radius of curvature at $p$ (which will be generically non zero). 

Within this region the metric will be approximately Minkowski, that is
\be
g_{ab}\,=\,\eta_{ab}+ \mathcal{O}(\e^2),
\ee
the order of approximation being fixed by the local curvature.
Finally, the local patch can be described via Riemann normal coordinates $\{x^a\}$, such that $p$ stays at $x^a=0$.  
\begin{figure}[h!]\label{fig:sptm}
\includegraphics[width=\columnwidth]{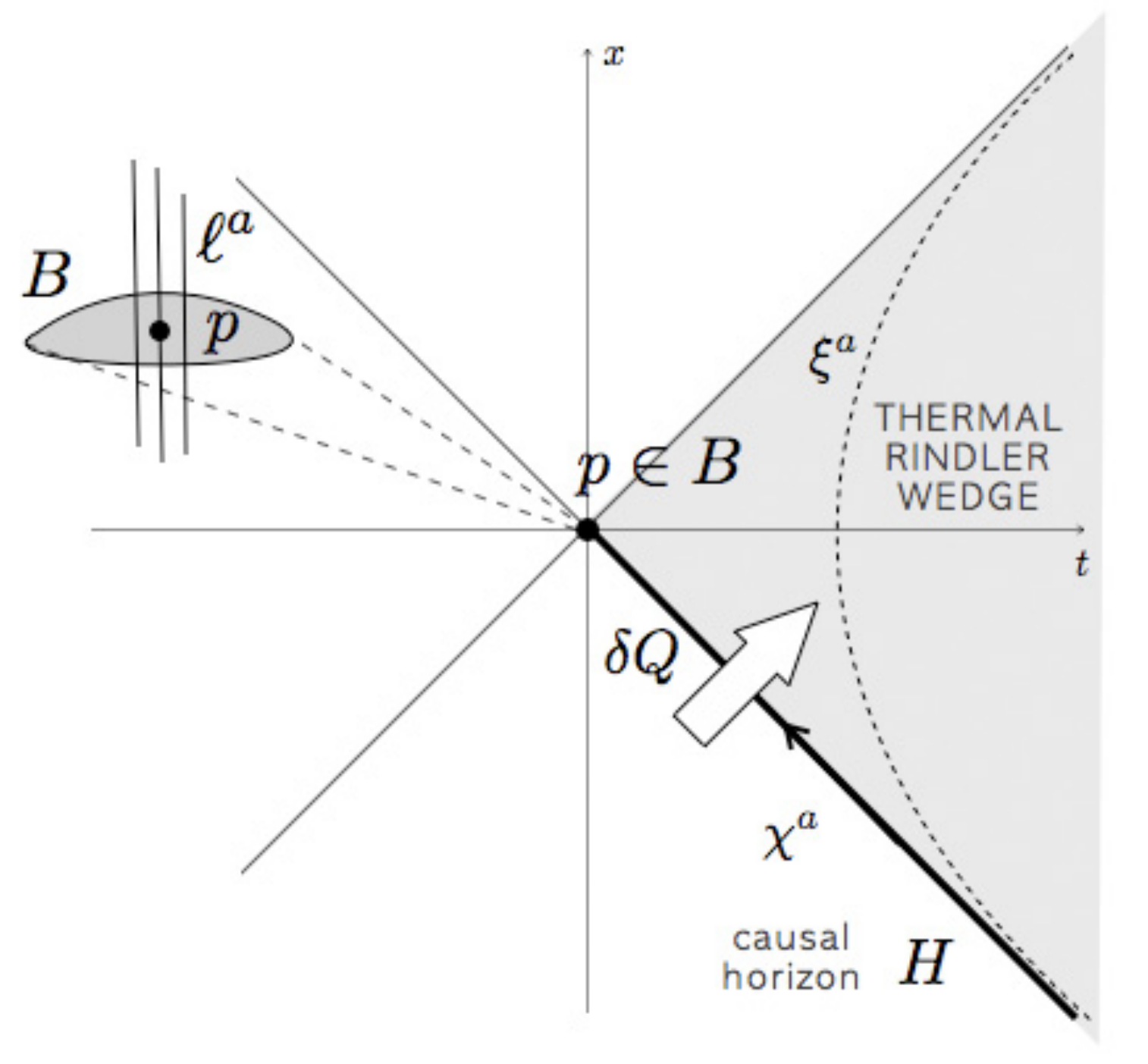}
\caption{A schematic view of the thermal Rindler wedge. The local causal horizon $H$ plays the role of a diathermic barrier for the thermal Rindler wedge. The thermal system is constituted by the set of Rindler observer $\xi^a$ moving along their isometry trajectories. For each of them the ensemble is described by $\rho=Z^{-1}exp(-H_b /T)$ and perturbed by the energy $\d Q$ that flows across $H$. }
\end{figure}
On the introduced LIF, one can construct a local Rindler frame (LRF) by the usual coordinate transformations.\footnote{Note that the construction of the LRF requires to fully use the equivalence principle, at least in its Einstein formulation, in order to identify geodesics motion~\cite{thom}.}
It is convenient here to choose $x=\, \chi \cosh (\eta \kappa)$ and  $t=\, \chi \sinh (\eta \kappa)$, introducing an arbitrary rescaling factor $\kappa$ for the proper time, in order to have a clear label for the Rindler wedge temperature in the following derivation.

With this choice, in the neighborhood of $p$, the LRF metric will be
\be
ds^2\,=-{\kappa^2}\chi^2 d \eta^2+ d\chi^2 + dy^2 + dz^2,
\ee
corresponding to the action of a Lorentz boost, with acceleration $a= 1/\chi$.

Therefore, within a small neighborhood of $p$, one can associate the boundary of the past of the patch $B$ to a section of the approximate Killing horizon, centered in $p$. 
The future pointing approximate boost Killing vector $\chi^a$, vanishing in $p$ by definition, will be tangent to the null congruence comprising the causal horizon and will leave invariant the tangent plane to $B$ at $p$.  

In terms of the approximate Killing vector $\chi^a$, one can introduce a time\footnote{Notice that there is no relation between the proper time $\eta$ defined in the wedge and the Killing parameter $v$ along the horizon. Nevertheless we need to keep the same scaling $\kappa$ for dimensional consistence. } label $v$ along the horizon null hypersurface, defined by $\chi^a\nabla_a  v=1$. The Killing parameter $v$ can be expressed in terms of the null congruence affine parameter $\lambda$. For a Killing horizon, the relation is generally given by 
\be
\lambda=-e^{-\kappa v},
\ee
so that the point $p$ is located at infinite Killing parameter and at $\lambda=0$.

As a consequence, one gets $\chi^a\,=\,(d\lambda/dv)\, \ell^a$, with $(d\lambda/dv)=-\kappa \lambda$ and, in the same way, 
\be
\hat{\theta}=\left(\frac{d\lambda}{dv}\right)\theta=- \kappa\lambda\, \theta \,\,\,\,\, \mbox{and}\,\,\,\,\,  \hat{\sigma}=\left(\frac{d\lambda}{dv}\right)\sigma=- \kappa\lambda\,  \sigma, \label{si} 
\ee
which gives some helpful relations between the Killing expansion $\hat{\theta}$ and shear $\hat{\sigma}$ and the respective affine geodesics quantities.

\subsection{Rindler Wedge Thermal Character} \label{trw}

Together with local flatness, the previous construction actually needs to further assume that the ground state of the fields living in the original spacetime is locally approximated by the Minkowski vacuum. In this case, with respect to the approximate Killing vector flow, associated with the set of observers living in the wedge at $x> |t|$ (Rindler wedge), the vacuum state can be interpreted as an approximate thermal state, with a temperature
\be
T \,\approx\,T_{un}\sqrt{-g_{00}}=\frac{\hbar \kappa}{2\pi}, \label{tw}
\ee
where $T_{un}=\hbar a/2\pi$ is the the Unruh temperature \cite{Un} for the Rindler observer. 

The expression in (\ref{tw}) shows that $T$ stays constant throughout the Rindler wedge, because of the gravitational Doppler factor $\chi$ associated with the Unruh temperature, and it is well defined on the horizon. Therefore, the thermal character of the Rindler state is effectively extended from the single Rindler observer to the whole wedge.  

As a further step, one can introduce an entropy for the system, via an entanglement argument. In the Rindler wedge, an accelerated observer can only access information on spacelike slices bounded by the bifurcation plane. Thereby, since vacuum fluctuation between the inside and the outside of the wedge are correlated, she/he will perceive an entanglement entropy, which scales with the area of the local boundary and diverges with the density of field states in the UV limit.
However, with the introduction of an UV cut-off (generically justified via the Planck scale quantum vibrations, the so called horizon \emph{zitterbewegung}) one can make this entropy become actually proportional to the area, that is
\be
S=\a A,
\ee
where the proportionality factor $\a$ can {\em a priori} depend on the nature of the quantum fields as well as be some complicate function of the position in spacetime \cite{'tHooft:1984re,Bombelli:1986rw}. Together with the temperature $T$, this notion of entropy allows to consider the local Rindler wedge with its Killing horizon as an analogue of a canonical ensemble (Gibbs state) bounded by a diathermic wall.

\subsection{Local Equilibrium Condition}   
Since all the thermal information related to the Rindler wedge is somehow recorded on the horizon boundary~\cite{Eling:2008}, one can define a notion of thermal equilibrium for the system in terms of the stationarity properties of the horizon.\footnote{It can be proved formally that the stability of the Rindler metric (Lorentz invariance), hence the Rindler horizon stationarity, actually implies the KMS conditions \cite{haa}, responsible for the thermal character of the vacuum energy fluctuations as measured by a Rindler observer.}
 
In this sense, the equilibrium state for the system is identified with the horizon cross-section at $p$, where the Killing expansion $\hat{\theta}$ and shear $\hat{\sigma}$ vanish and the horizon is instantaneously stationary with respect to $\lambda$,  
\be
dS=\d(\alpha  A)=0.
\ee
Given the relations in (\ref{si}), it is clear that the equilibrium state can be reached as long as the affine quantities $(\theta,\sigma)$ are not diverging in $p$.  
The local curvature of $B$ at $p$ determines the geometrical properties of the horizon null congruence and consequently the value of affine expansion and shear at $p$. Hence, in order to define an equilibrium surface one just need to require a suitably smooth curvature for $B$ at $p$, without fixing {\em a priori} the values of affine expansion and shear. 

This is a very delicate point, as we will see that the properties of $B$ at $p$, and the corresponding values of the optical scalars of the associated null congruence, actually select the theories of gravity which may arise from the thermodynamical approach by fixing the gravitational degrees of freedom of the theory.
In this sense, we will show that the choice of $B$ can be directly related to the Equivalence Principle formulation, which plays a fundamental role in the argument.

\section{Review of the Previous Results: the GR case} \label{ees}

Moving away from the bifurcation surface at $p$, the spacetime will become dynamical and the presence of matter will distort the local Rindler causal structure, perturbing the thermal state at the same time. 

As long as the departure from equilibrium is small and slow, it can be described in analogy with a quasi-static process where a suitably small amount of energy is thrown through the horizon. Then, one needs some condition to relate the spacetime geometry deformation to the variation of the fields energy content. In fact, thanks to the thermal properties of the system, this condition can actually be borrowed from classical equilibrium thermodynamics.

Indeed, for a slightly perturbed Gibbs state at temperature $T$, variations of entropy and internal energy are related by the Clausius relation,
\be
dS=\frac{\d Q}{T}\label{cl}\, ,
\ee
where the change in the mean energy is taken to be due to the fluxes into/from an unobservable region of spacetime, which is perfectly thermalized by the horizon system, and hence it is assumed to consist entirely of heat. For a system undergoing a quasi-static process of energy exchange with the surroundings, relation (\ref{cl}) is nothing but a local thermodynamical equilibrium condition. 

Now, since the thermal Rindler wedge behaves like a Gibbs state, the first fundamental assumption in \cite{Jacobson:1995ab} consists in using relation (\ref{cl}) to relate the horizon entropy and the boost energy across the Rindler horizon, with $T$ given by eq.~\eqref{tw}.

As a second fundamental postulate, it is assumed a universal entropy density $\alpha$ per unit horizon area $\delta A$, such that 
\be
d S= \alpha \delta A, \label{ent}   
\ee
thereby implicitly considering a constant UV cut-off for the fields, whereas in general one would have $dS=\d(\a A)$ (i.e.~$\alpha$ can be some spacetime function).
With this choice, the changes of the entanglement entropy of the fields in the wedge can be effectively described in terms of geometrical variations of the horizon cross-section.

Let us highlight here that the $\alpha=\mbox{constant}$ assumption, made in the GR derivation of \cite{Jacobson:1995ab,Eling:2006}, can be indeed recast as an explicit choice of a specific formulation of the Equivalence principle. As said, generally the UV cut-off $\alpha$ is fixed at the quantum gravity scale. This can be identified as the scale at which the gravitational action is of the order of the quantum of action $\hbar$. For GR this is the standard Planck length $l_p=\sqrt{G\hbar/c^3}$ and hence it is directly related to the Newton constant. 
However, for a general scalar-tensor theory, i.e.~a theory compatible just with the Einstein Equivalence principle (EEP)~\cite{Willl}, $G$ is promoted to a spacetime field. As a consequence of this, one should expect that the cut-off will be generically position dependent. In this sense, assuming $d S= \alpha \delta A$ is equivalent to assume the strong formulation of the Equivalence Principle (SEP)~\cite{Willl}, hence effectively to allow for the only two SEP-compatible gravity theories: GR and Nordstr\"om gravity~\cite{Gerard:2006ia}.

Given the assumption in (\ref{ent}), a quantitative expression for the system entropy variation is obtained just by applying the definition for the change of the horizon area in terms of the expansion rate of the null geodesics comprising it, that is
\be
\delta A= \int_{H}\vS\,  \theta \,d\lambda \label{arria}\,, 
\ee
with $\vS$ indicating the 2-surface area element of the horizon cross-section.
Moving away from the equilibrium surface at $\lambda=0$, along the null congruence, the infinitesimal evolution of $\theta$ is given by a linear expansion around its equilibrium value at $p$, up to the first order in $\lambda$,
\be
\theta \approx \theta_p+\lambda \, \left. \frac{d\theta}{d \lambda} \right|_p\,+\mathcal{O}(\lambda^2) \label{exp}\,. 
\ee
This first order coefficient will be determined as usual by the Raychaudhuri equation,
\be
\frac{d\theta}{d \lambda}= -\frac{1}{2}\theta^2-\|\sigma\|^2-R_{ab}\ell^a\ell^b \label{Ray}\, , 
\ee
where $\|\sigma\|^2$ stands for the squared congruence shear $\sigma^{ab}\sigma_{ab}$.\footnote{Here we consider a vanishing twist, as the null congruence is taken hypersurface orthogonal.}

In this way, the entropy variation, up to $\mathcal{O}(\lambda^2)$, is given by 
\be 
d S= \alpha  \int_{H}  \vS \, d\lambda\, { \left[\theta -\lambda \left(\frac{1}{2}\theta^2+\norm\sigma\norm^2+R_{ab}\ell^a\ell^b\right) \right]}_p \label{ee}\, .
\ee

The mean energy variation of the thermal system, given by the boosted energy current flux of matter, is then described by the heat flux across the horizon as
\be 
\delta Q=  \int_{H}  T_{ab}\chi^a d\Sigma^b \label{hea} \, ,
\ee
where $T_{ab}$ is the matter stress energy tensor, while the volume element is given by $d\Sigma^b=\vS \, d\lambda \,\ell^b$. The same quantity, with respect to the null congruence parameters, reads
\be 
\delta Q=  \int_{H} \vS \, d\lambda \,\, (-\lambda \kappa)\,\,T_{ab}\ell^a\ell^b  \label{heaa}\, . 
\ee

At this point, asking for relation (\ref{cl}) to hold for all null vectors $\ell^a$, one can equate the $\mathcal{O}(\lambda)$ integrands in (\ref{ee}) and (\ref{heaa}).
At the zeroth order in $\l$, the value of heat flux at $p$ is zero, hence one necessarily gets $\theta_p=0$. Then, to the first order, 
\be 
\frac{2\pi}{\hbar \a}\, T_{ab} \ell^a \ell^b= (\|\sigma\|^2+R_{ab}\ell^a\ell^b)_p \label{eoss}\, .
\ee
Moreover, if one further requires that $\s_p=0$, one is left with
\be
\frac{2\pi}{\hbar \alpha}\,T_{ab}= R_{ab}\,+\,\Phi\,g_{ab} \label{eos0} \, ,
\ee
where $\Phi$ is an undetermined integration function. 

Eventually, by assuming the local energy conservation, that is $\nabla^b T_{ab}=0$, applying the divergence operator on both sides of (\ref{eos0}), and using the contracted Bianchi identity $\nabla^bR_{ab}=\frac{1}{2}\nabla_a R$, one finally gets $\Phi=-\frac{1}{2}R-\Lambda$, hence
\be
\frac{2\pi}{\hbar \alpha} \,T_{ab}= R_{ab}\,-\frac{1}{2}R\,g_{ab}\, -\,\Lambda\, g_{ab}\, , \label{eos1}
\ee
where $\Lambda$ is some arbitrary integration constant.
Once the condition
\be
\alpha=\frac{1}{4 \hbar G} \label{alfa}
\ee
is imposed, one can easily recognize the familiar Einstein equations. Noticeably, eq.~\eqref{alfa} implies that the entropy density of the local Rindler horizon is the same as the one of a black hole. 

This result tells us that, given the entropy and energy conservation for the system, the local thermodynamical equilibrium condition is, in fact, equivalent to the Einstein equation for the local thermal spacetime. Furthermore, the EP implies that the above construction can be done at {\em any} spacetime point $p$, and hence that equation \eqref{eos1} holds everywhere in spacetime.

On the other hand, by allowing for some shear in $p$, that is for $\s_p\ne0$, the Einstein equation is no more recovered. As discussed in the previous section, the LRF is not sensitive to the exact value of the affine expansion and shear at $p$. Therefore, setting $\sigma_p =0$ is an unjustified arbitrary choice.  This was firstly realized in~\cite{Eling:2006}. In fact, a non zero affine shear at $p$ may change the way in which the equilibrium is approached by the system in the Killing frame. Given the relations in (\ref{si}) and the analogue of the expansion (\ref{exp}) for the affine shear, one realizes that the Killing shear falls off to zero at $p$ as $\hat{\s}\sim e^{-2\k v}$ when $\s$ vanishes, while only as $\hat{\s}\sim e^{-\k v}$ when $\s$ is non vanishing. In this sense, for a non vanishing affine shear, the equilibrium approach can be considered slow enough for the system to be in a non-equilibrium regime. 

This argument was used in \cite{Eling:2006} to recast the thermodynamical derivation in a non-equilibrium setting, where 
\be dS>\d Q/T.\label{sec}\ee
In this new context, the Clausius law is replaced by the entropy balance law,
\be
dS=\d Q/T+d_i S, \label{ebal}
\ee
and the extra shear term in (\ref{eoss}) is then associated with the internal entropy production $d_iS$, generated by the system out of equilibrium. The internal entropy contribution, $\mathcal{O}(\l)$, has the form
\be
d_iS= - \frac{4\pi \eta}{\hbar}\int_{H} \vS \, \l\, d\l\, \norm\sigma\norm^2_p\,  \label{verintent}
\ee
and, in analogy to the internal entropy production terms originating from the squared gradients of state variables, a universal property of systems with viscosity in non-equilibrium thermodynamics \cite{deGroot}, it is interpreted as an internal entropy production term due to some internal spacetime viscosity, with $\eta=\hbar \a/ 4 \pi$. Furthermore, it is also noticed in \cite{Eling:2006} that by using \eqref{alfa}, one gets $\eta=1/(16\pi G)$ in agreement with the value obtained for the shear viscosity of the stretched horizon of a black hole in the so called membrane paradigm  \cite{Price:1986yy,membrane}. This result concludes the review of the argument described in \cite{Jacobson:1995ab, Eling:2006, Eling:2008}.\\
 
\subsection{Some Remarks}

One can consider the local spacetime thermodynamics developed in \cite{Jacobson:1995ab, Eling:2006, Eling:2008} as a useful experimental setting, where the local behaviour of gravity is tested in its different formulations.  Nevertheless, we believe that the argument  still misses a clear understanding.

First of all, it seems that the thermodynamical approach cannot have a detailed control over the degrees of freedom of the resulting gravity theory. In particular, on the way they are effectively involved in the perturbation of the spacetime causal structure.  Indeed, by allowing for some shear at $p$, the local equilibrium condition is formally split in two parts: the $\ell^a$ part, which leads to the Einstein equation, and the $\partial \ell^a$ part, related to the shear term, which remains unexplained. In this sense, at the $\partial \ell^a$ level, the local equilibrium condition is broken.

Therefore, equilibrium thermodynamics can give a suitable description only under the assumption that the affine congruence orthogonal to $B$ has zero expansion and shear at $p$. However, we saw that  this is equivalent to require that the chosen $B$ (and hence its associated null congruence) is less general than the one allowed by the assumed entropy-area relation (or alternatively by the SEP).  Such an ansatz seems too restrictive to consider (\ref{eos1}) as a general result.

On the one hand, allowing for non zero affine shear at $p$ give rise to further interesting clues. In some way, the shear contribution  in (\ref{eoss}) brings into the entropy balance process a new degree of freedom, which is not fixed by the Ricci tensor and so has nothing to do with the local matter energy sources. Actually, the surface shear is generally related to the Weyl tensor and usually associated with the distortion on the geodesics congruence due to a gravitational perturbation.

In fact, this argument opens an issue about the absence so far of any role for gravitational fluxes in the system energy perturbation mechanism. Due to their non-local nature, the gravitational energy fluxes cannot be taken into account with a proper stress-energy tensor (SET). However, allowing for non-local terms, as the one in $\|\sigma\|^2$ in (\ref{eoss}), seems at odds with neglecting the role of these non-local energy contributions.

On the other hand, the interpretation of the internal entropy contribution as a by-product of some sort of viscous work on the system, given in \cite{Eling:2006, Eling:2008}, is very reasonable, because the term (\ref{verintent}) is actually related to some mechanical deformation due to the presence of shear in the null congruence generating the horizon. 

In this sense, such a spacetime viscosity seems naturally related to the distorsive effect of a gravitational flux, to be intended as a local curvature perturbation which is independent from the Einstein equation. This suggest that gravitational energy fluxes can possibly play a role into the total entropy balance of the system without entering into the Einstein equilibrium relation.

Starting from these remarks, we are led to reconsider the thermodynamical argument of this section in a fully non-equilibrium setting further elaborating on the approach taken in \cite{Eling:2006}.
In particular, for the motivations given above, we argue that the internal entropy production, such as (\ref{verintent}), actually indicates the presence of dissipative (irreversible) processes, to be related to the conformal components of the spacetime curvature, which do not take part into the field equations and are associated with purely gravitational degrees of freedom.

\section{Internal entropy in non-equilibrium thermodynamics}\label{ne-th}

In classical non-equilibrium thermodynamics, the rate of change of the entropy is generally written as the sum of two contributions:
\be
dS=d_eS+d_iS, \label{ento}
\ee
where $d_eS$ is the rate of entropy exchange with the surroundings while $d_iS$ comes from the process occurring inside the system and is a non-negative quantity, accordingly to the second law of thermodynamics. In particular, $d_iS$ is zero for reversible (quasi-static) processes and positive for irreversible processes.

The Clausius relation used for the equilibrium approach in section III, is actually equivalent to the Clausius definition of entropy for the equilibrium system, that is
\be
d_eS=\delta Q/T \,\,\,\,\, \mbox{and}\,\,\,\,\, d_iS=0,
\ee
as, in that case, the horizon perturbation is effectively described as a quasi-static process occurring in continuous equilibrium with the surrounding. 
However, this definition does not hold true any more as irreversible processes come into play. 

Actually, in the non-equilibrium thermodynamical setting, the Clausius definition of entropy is generalized to the expression
\be
d S\,=\, \frac{\delta Q}{T}\,+\, \d N, \label{unhe}
\ee
where $\d Q$ is classically referred to as \emph{compensated heat\,}, that is the heat transferred between the system and its surroundings, while $\delta N$, the so called \emph{uncompensated heat\,}, indicates the amount of entropy associated with the heat which is intrinsic to the system when it undergoes an irreversible process.

Let us stress that the above definition is very general, as it does not require either an a priori specification of the nature of the non-equilibrium variable, or the nature of the process involved. It generalizes the notion of local equilibrium by extending the entropy balance to the unbalanced contributions related to the irreversible processes, like dissipation (see e.g.~\cite{cit}).

The generalized Clausius relation  (\ref{unhe}) is helpful in order to clarify the nature of the equilibrium and non-equilibrium contributions defining our system entropy. In fact, by using the definition of the non-equilibrium entropy given in (\ref{ento}), we can write
\be
d_eS\,+\,d_iS\,=\, \frac{\delta Q}{T}\,+\, \d N \, , \label{ccc}
\ee
and identify the external and internal entropy in terms of the compensated and uncompensated heat, respectively
\begin{itemize} 
\item $d_eS \,= \,\d Q/T$, at the reversible level,
\item $d_iS \,= \,\d N$, at the irreversible level.
\end{itemize}
With this approach, the argument described in \ref{ees} acquires a straightforward interpretation. 
Indeed, the extra contribution (\ref{verintent}) introduced by the non vanishing horizon shear is an internal entropy production term allowed by the most general choice of the null congruence associated with B compatible with the area-entropy relation for GR (that we linked to the choice of the EP formulation). Therefore, it has to be seen as a by-product of the presence of internal/purely gravitational degrees of freedom of the theory which can be responsible for irreversible dissipative processes.

However, in order to physically identify an internal entropy contribution $d_iS$ into the general expression for the horizon entropy given in (\ref{ee}), one needs a clear understanding of the relation between non-equilibrium forces and intrinsic spacetimes properties involved.  

Since all the thermal information of the Rindler wedge vacuum is recorded on the horizon boundary \cite{Eling:2008}, the internal spacetime variables involved in the non-equilibrium process should be related to the null geodesic congruence kinematics around $p$. In this sense, a possible way to capture non-equilibrium features of the thermal system is to use the analogy between the congruence bundle comprising the horizon and a classical fluid.

\subsection{Null Congruence as a Continuous Medium} \label{nccm}

Given the Price and Thorne \emph{membrane} approach~ \cite{Price:1986yy,membrane}, the local Rindler horizon can be effectively approximated by a timelike hypersurface living just inside the true Rindler wedge, i.e.~a timelike \emph{stretched horizon}.

With a 2+1 decomposition, the timelike congruence comprising the stretched horizon is formally equivalent to a 2+1 continuous medium (fluid) living on the spacelike two dimensional cross section of the hypersurface and moving with velocity $\bold{v}^i$, defined by the unit timelike vector tangent to the hypersurface \cite{Price:1986yy, Eling:2008}.

As the stretched horizon tends to the Rindler horizon, the unit timelike vector $\bold{v}^i$ approximates the Killing vector $\chi^a$. Therefore, from a purely kinematical point of view, 
we can consider the velocity gradient of the medium, $\nabla_j \bold{v}_i$, as the equivalent of the deformation tensor field of our congruence (extrinsic curvature), in the Killing frame,
\be
\hat{B}_{ab}={h_a}^c {h_b}^e\chi_{c;e}, 
\ee
where the tensor $h_{ab}$ is the projector onto the spacelike cross-section transverse to the congruence (transverse metric), given by 
\be
h_{ab}=g_{ab}\,+\,\chi_a\, k_b\,+\,\chi_b\, k_a
\ee
while $\chi^a$ is the Killing vector flow, as defined in section II and $k$ is an auxiliary null vector field satisfying $k^a\,\chi_a=-1$ which spans, together with $\chi^a$ the 2-surface horizon cross section. 
In both cases, the tensor can be decomposed into trace and symmetric trace-free part,
\be
\bold{V}_{ij}=\frac{1}{3}(\nabla^\rho  \bold{v}_\rho) \bold{\delta}_{ij}\,+\,\tilde{ \bold{V}}_{ij}, \label{V}
\ee
where $\bold{V}_{ij}$ represents the symmetric component of $\nabla_j \bold{v}_i$, $\bold{\delta}_{ij}$ is the identity tensor and 
\be
\tilde{ \bold{V}}_{ij}=\frac{1}{2}(\nabla_i  \bold{v}_j\,+\,\nabla_j  \bold{v}_i)
\ee
is the deviatoric traceless tensor. Similarly
\be
\hat{B}_{ab}=\frac{1}{2}\hat{\theta} h_{ab} \,+\,\hat{\sigma}_{ab}, \label{B}
\ee
where $\hat{\theta}$ and $\hat{\sigma}_{ab}$, are the usual scalar expansion and shear of the null congruence. 

Now, for the continuous medium, the presence of $\bold{V}_{ik}$ indicates a relative motion between various parts of the fluid and it is responsible for the irreversible viscous transfer of momentum into the medium and for the consequent production of internal entropy. In particular, for the viscous medium, the mechanical dissipation will contribute to the internal entropy density $d_is$ by an amount\footnote{Among the several dissipative processes which can contribute to the internal entropy production for the medium, here we just consider the viscous stresses.}
\begin{eqnarray}
d_is=- \frac{1}{T} \bold{P}_{ij}^{\rm vis}\,\bold{V}^{ij}\,=-\frac{1}{T} \,(\bold{p}^{\rm vis}\,\nabla^\rho  \bold{v}_\rho+\tilde{\bold{P}}_{ij}^{\rm vis} \tilde{ \bold{V}}^{ij}),\label{coco}
\end{eqnarray}
where $\bold{P}_{ij}^{\rm vis}=\bold{p}^{\rm vis}\,\bold{\delta}_{ij}+\tilde{\bold{P}}_{ij}^{\rm vis}$ is the viscous pressure acting on the medium, once decomposed, respectively, in its bulk and trace-less components \cite{cit}. 

%%%%%

%%%%%

\subsection{Constitutive equations}\label{subsec:const-eq}

The expression for the internal entropy in (\ref{coco}) is a bilinear form, where the external stress $\bold{P}^{\rm vis}$ comes coupled with its conjugate strain $\bold{V}$. Typically, $\bold{V}$ is a known function of the internal state variable of the medium, while the external forces are unknown. However, in classical irreversible thermodynamics, under a \emph{local equilibrium hypothesis}, that is, as far as it is possible to assume that local and instantaneous relations between the thermal and mechanical properties of the system are the same as for uniform systems at equilibrium, the deformation of the velocity field and the viscous pressure (mechanical stress) can be related by the linear constitutive equations, 
\begin{eqnarray}
\bold{p}^{\rm vis}&=&\,-\,\zeta\,\nabla^\rho  \bold{v}_\rho\,,\qquad\mbox{Stoke's law}\label{eq:co1}\\
\tilde{ \bold{P}}_{ij}^{\rm vis}&=&\,-\,2\eta\, \tilde{ \bold{V}}_{ij}\,,\qquad\mbox{Newton's law}\label{eq:co2}
\end{eqnarray}
where $\zeta$ and $\eta$ respectively indicates the bulk and shear viscosity of the medium. The dissipation coefficients are exactly related to the time dependence of equilibrium fluctuations for $\bold{p}^{\rm vis}$ and $\tilde{ \bold{P}}_{ij}^{\rm vis}$, in the framework of the linear response theory, via the Green--Kubo relations \cite{cit}.

Thereby, given the linear relation introduced above, the viscous internal entropy density can be defined in terms of velocity gradient components of the medium, that is
\begin{eqnarray}
d_is= \frac{\zeta}{T}\, (\nabla^\rho \bold{v}_\rho)^2\,+\,\frac{2\eta}{T} \, {\norm{\tilde{\bold{V}}}\norm}^2 ,\label{fee}
\end{eqnarray}
where ${\norm{\tilde{\bold{V}}}\norm}^2=\tilde{ \bold{V}}_{ij}\tilde{ \bold{V}}^{ij}$.

Now, given the stretched horizon approximation and by associating the bulk term $(\nabla^\rho  \bold{v}_\rho)$ to $\hat{\theta}$ and the deviatoric traceless tensor $\tilde{ \bold{V}_{ij}}$ to $\hat{\sigma}_{ab}$, one can expect a dissipative internal entropy production term for the horizon congruence, of the form
\begin{eqnarray}
(d_iS)^{\rm vis}=\frac{1}{T}\int_H\,\vS \,dv \,\zeta\,\hat{\theta}^2\,+\,2\eta\,{\norm{\hat{\sigma}}\norm}^2. \label{eee}
\end{eqnarray}

The above expression provides two important insights. First, it identifies the congruence kinematical quantities which are responsible for the irreversible viscous transfer of momentum into the system, and for the consequent production of internal entropy. In our setting, the favorite candidate to play the role of conjugate viscous stresses are the spacetime curvature perturbations. Indeed, the evolution of the deformation tensor $\hat{B}_{ab}$ along the approximated Killing flow is driven by the equation
\be
\chi^c\nabla_c \hat{B}_{ab}\,=\,\kappa\hat{B}_{ab}\,-\,\hat{B}_a^c\hat{B}_{cb}\,-\, {h_a}^c {h_b}^d R_{cedf}\chi^e\chi^f, \label{db}
\ee
whose trace and trace-free part give respectively the Raychaudhuri and shear evolution (tidal force) equations along the horizon~\cite{poibook}. In vacuum, the only source term for both the equations is given by the electric part of the Weyl tensor, $C_{ab}={h_a}^c {h_b}^e C_{cdef} \chi^d  \chi^f$ \cite{poibook,membrane} which is well known to be associated with gravitational perturbations. 

Secondly, the strict correspondence between equations \eqref{fee} and \eqref{eee}, seems to suggest the presence of linear constitutive equations, analogues of \eqref{eq:co1},\eqref{eq:co2}), between the horizon kinematical quantities $\hat{\theta}$ and $\hat{\sigma}_{ab}$ and some external viscous stress. Therefore, by extending the fluid analogy argument, one should be able to relate the congruence viscosity coefficients to the local curvature fluctuations via some analogue of the Green--Kubo relations. We leave this for future investigations.

\section{Non-equilibrium Spacetime Thermodynamics: GR} \label{neos}

We now have a clear way to relate the non-equilibrium features, arising in the thermodynamical derivation of the Einstein equations in section \ref{ees}, to those kinematical degrees of freedom of the horizon congruence which are turned on by the local spacetime curvature.  The horizon kinematics actually defines the intrinsic spacetime properties involved in the irreversible processes.

We can then reproduce the thermodynamical derivation of the Einstein equations with a non-equilibrium irreversible thermodynamics setting, simply by starting, in a quite general way, from a generic spacelike 2-surface patch $B$ at $p$, with non vanishing $\theta_p$ and $\sigma_p$.  

As for the previous derivation, the strong equivalence principle allows us to use the entropy area relation as in (\ref{ent}), that is
\be
d S\,=\,\alpha \d A \,\,\,\,\,\mbox{with}\,\,\,\,\,\alpha=\mbox{const}.
\ee
However, in the new setting, the Clausius relation will be generalized to the expression \eqref{unhe}, 
where $\d N$ will now encode all the information about both microscopic properties  and  irreversible perturbations of the system. 

The new argument starts from the same definition of entropy, given in (\ref{ee}),
\be
d S= \alpha  \int_{H}  \vS \, d\lambda\, {\left[\theta -\lambda \left(\frac{1}{2}\theta^2+\|\sigma\|^2+R_{ab}\ell^a\ell^b \right) \right]}_p. \label{Ks}
\ee
Since we are now dealing with a non-equilibrium setting, we expect that the entropy can be expressed as a sum of two different contributions $dS=d_eS+d_iS$. Moreover, for the argument given in \ref{nccm}, we are able to identify the form of the non-equilibrium, unbalanced, entropy terms. Therefore, we can write
\begin{eqnarray}
d_e S &=&\alpha  \int_{H}  \vS \, d\lambda\, (\theta -\lambda R_{ab}\ell^a\ell^b )_p\, \label{rev} \\
d_i S &=& - \alpha  \int_{H}  \vS \, d\lambda\,\lambda{\left( \frac{1}{2}\theta^2+\|\sigma\|^2 \right)}_p\, ,  \label{irrev} 
\end{eqnarray}
and separate, as previously argued, the reversible and irreversible levels
\begin{itemize} 
\item \eqref{rev} $= \,\d Q/T$, at the reversible level,
\item \eqref{irrev} $= \,\d N$, at the irreversible level.
\end{itemize}

From the first expression above, one has
\begin{eqnarray} 
d_eS \,=\alpha  \int_{H}  \vS \, d\lambda\, \left(\theta -\lambda R_{ab}\ell^a\ell^b \right)_p&=&\\ \nn
=- \frac{2\pi}{\hbar}\int_{H} \vS \, d\lambda \,\lambda \,T_{ab}\ell^a\ell^b &=& \frac{\d Q}{T},
\end{eqnarray} 
where the heat flux is still defined by the expression in (\ref{heaa}). Even for the non-equilibrium setting the reversible heat  will vanish at $\lambda=0$. Thereby, at the zero order in $\lambda$, one deduces again $\theta_p=0$, while, at the first order, the relation $R_{ab} +\,\Phi g_{ab} = (2\pi/\hbar\alpha)T_{ab}$, is recovered for all null vectors $\ell^a$. Following the previous discussion this implies, together with the conservation of the matter stress-energy tensor, the Einstein equations if $\alpha=(4\hbar G)^{-1}$.

On the other hand, for the irreversible level, we have, in accordance with \eqref{verintent},
\be
\delta N\,=\,d_iS= \,- \,\alpha\, \int_H \vS\,d\lambda\, \lambda \, \|\sigma\|_p^2 \, . \label{ieppo}
\ee
This again identifies the shear contribution as an internal entropy term, associating it to some irreversible dissipative process occurring in the thermal Rindler wedge.

To get a physical interpretation of $\d N$ with respect to the thermal properties of the Rindler wedge, it is helpful to express equation (\ref{ieppo}) in terms of the Killing horizon parameters. In the new frame,
\be
\delta N\,=\,d_iS= \, \frac{\alpha}{\kappa} \int_H \vS \,dv \,  \|\hat{\sigma}\|_p^2 \ge 0, \label{Kieppo}
\ee
in accordance with the second law of thermodynamics.

By a comparison with expression (\ref{eee}), one can  actually interpret the expression in (\ref{Kieppo}) as the standard entropy production term for a fluid with shear viscosity $\eta$, defined by 
\be
\frac{2\eta}{T}=\frac{\alpha}{\kappa}, \label{relo}
\ee
that is $\eta=\hbar \a/4\pi$, in agreement with  the universal relation for the viscosity to entropy density ratio found in the AdS/CFT context~\cite{malda}.

While the previous discussion shows that the spacetime thermodynamics nicely fits into a non-equilibrium setting, we now want to take this arguments a step further and ask whether the expression in (\ref{Kieppo}) can be effectively related to some gravitational energy flux.

%\subsection{Tensorial Degrees of Freedom and Gravitational Dissipation}

The expression for the \emph{uncompensated heat} given in (\ref{Kieppo}) quantifies the energy of the system which is effectively dissipated by the viscous process,
\be
T\,\d N=\frac{\a \,T}{\kappa}\,\int_H \vS \,dv \,  \|\hat{\sigma}\|_p^2. \label{gflu}
\ee
Then, by substituting $\alpha=(4\hbar G)^{-1}$, from the reversible sector of the thermodynamical approach, the quantity in (\ref{gflu}) reads
\be
T\,\d N= \frac{1}{8 \pi G} \int_H \vS \,dv \,  \|\hat{\sigma}\|_p^2, \label{gfflu}
\ee
which coincides with the Hartle-Hawking formula for the tidal heating of a classical black hole~\cite{TP, HH, Chandra, Poi:2004, Poi:2005}.\footnote{Note that both in \cite{TP} and in \cite{Chandra} the Hartle-Hawking formula expressing the relation between the horizon area variation and the horizon shear is utilized with a definition of the surface gravity $\epsilon$ which is half of that used in \cite{HH}.}  

This is a striking result as it defines the internal entropy production as a purely gravitational effect. 
Indeed, it can be associated with the work done on the horizon by the perturbative tidal field which is described by the electric part of the Weyl curvature tensor.~\footnote{Even though the magnetic part of the Weyl tensor is actually necessary in order to define the Weyl curvature and the equations governing its propagation (Bianchi identities), this part does not play any direct role in determining the time derivative (evolution) of the congruence kinematic quantities, as it is just related to their spatial gradients.} Furthermore, relation \eqref{relo} suggest that such a work has to be seen as acted upon the internal/microscopic degrees of freedom of the theory rather than on macroscopic quantities (in this sense \eqref{gfflu} cannot be interpreted as a standard/reversible work term). The horizon viscosity implies that such a work will be converted into internal heat. Hence, the presence of the internal entropy term can then be directly related to the process of dissipation via gravitational/internal degrees of freedom. In this sense, the irreversible sector contains the information about the possible activation/propagation of such degrees of freedom of the theory. 

\section{Non-equilibrium Spacetime Thermodynamics: $f(R)$}\label{sec:fR}

A crucial assumption in the previous derivation was the validity of the SEP which allowed to consider the entropy density $\alpha$ as a constant. One might wonder what are the consequences of relaxing such an assumption in favor of the less restrictive Einstein Equivalence Principle (EEP). In this case, one might generically expect that the entropy density is promoted to a spacetime function (basically because the EEP implies a spacetime dependent Newton constant). However, in the definition for the entanglement entropy of the Rindler wedge, this implies a possibly very complicated spacetime dependence for the UV cut-off. Furthermore, the specific form of such a cut-off is not uniquely fixed by the EEP correspondingly to the fact that the latter generically allows for many generalized theories of gravity. 

In order to make the argument as simple as possible, and following the treatment of \cite{Eling:2006}, we consider here the specific case of $f(R)$ gravity, which is known to be equivalent to a single field scalar-tensor theory (more precisely a Brans-Dicke theory with $\omega=0$ and a specific potential for the scalar field~\cite{Sotiriou:2008rp}). In this case, the UV cut-off is known to be proportional to some function of the curvature $\bold{f}(R)\equiv f'(R)$ (where the prime indicates the derivative with respect to $R$), playing the role of the inverse of the gravitational coupling. In this case, the area entropy relation is known to be given by
\be 
S=\a\,\bold{f}(R)\, \vS \label{coffy}
\ee
where $\a$ is still a constant (albeit {\em a priori} different from the one considered in the previous section). 

It is easy to see that in this case the entropy variation along the null congruence will be
\be 
\frac{dS}{d\lambda}=\a \left ( \,\frac{d\bold{f}}{d\lambda}\, \vS\,+\,\bold{f}\,\frac{d\,\vS}{d\lambda}\, \right),
\ee
where, by definition $\vS^{-1}d\,\vS/d\lambda=\theta$.  

Consequently, the entropy change along the horizon will read~\cite{Eling:2006}
\be 
dS= \alpha\, \int_H \vS\,d\lambda\, (\dot{\bold{f}}\,+\,\bold{f}\,\theta), \label{enst}
\ee
therefore acquiring, with respect to the previous argument, an extra contribution $\dot{\bold{f}}$ coupled to the dynamics of the scalar function $\bold{f}$. (Here the dot stays for differentiation with respect to $\lambda$.)

For this reason, in order to set the instantaneous stationarity condition at $p$, that is $dS=0$, the affine expansion is no more a good dynamical variable. In this sense, it is helpful to define the quantity $\tilde{\theta}\equiv(\theta\, \bold{f}\,+\,\dot{\bold{f}})$ as a sort of effective expansion for the congruence. Consequently, the equilibrium surface for the system will be fixed by the condition
\be 
\tilde{\theta}_p=0\, , \label{rela}
\ee
that is $\theta_p=-\dot{\bold{f}}/\bold{f}$, where $\dot{\bold{f}}=\bold{f}'(R)\,\ell^a\,{R,}_a$ is generally nonzero. 
In particular, this actually provides an example of LRF equilibrium surface, for which $\theta_p$ is always non-vanishing, apart from the trivial case where $\bold{f}$ is constant, for which the theory will be equivalent to GR.

From section \ref{nccm}, we could already expect that the presence of the non-vanishing affine expansion would produce a non-equilibrium contribution to the system entropy.
In order to get a quantitative expression for the entropy change in the neighborhood of $p$, one again can consider an infinitesimal deviation of the entropy from its equilibrium value. 

Let us then Taylor expand the integrand in (\ref{enst}) around $p$ up to the first order in $\lambda$, that is
\be 
\tilde{\theta}=\,\tilde{\theta}_p\,+\,\lambda\, \left. \dot{\tilde{\theta}} \right|_p\,+\,\mathcal{O}(\lambda^2),
\ee
where 
\be
\dot{\tilde{\theta}}_p\,=\,(\ddot {\bold{f}}\,-\,\bold{f}^{-1}\dot {\bold{f}}^2\,+\, \bold{f}\,\dot \theta)_p. \label{etr}
\ee

One can use the Raychaudhuri equation (\ref{Ray}) and the geodesic equation $\ell^a{\ell^b}_{;a}=0$, to obtain the $\mathcal{O}(\lambda)$ expression for the entropy change
\begin{eqnarray} 
dS=\alpha\, \int_H \vS\,d\lambda\, \lambda \,\left[ (\bold{f}_{;ab}\,-\,\bold{f}\,R_{ab})\,\ell^a\ell^b\,- \right.\\ \nn
\left.-\,3/2\,\bold{f}\,\theta^2\,-\,\bold{f}\,\norm\sigma\norm^2\right]_p,\label{eq:ds}
\end{eqnarray}
where relation (\ref{rela}) is used to substitute $\bold{f}^{-2}\dot {\bold{f}}^2=\theta^2$ at $p$.
Now, keeping the expression in (\ref{heaa}) for the heat flux, one can finally reproduce the same approach used in section \ref{neos}. 

At the reversible level, the generalized Clausius relation gives 
\be
\bold{f}\,R_{ab}\,-\,\bold{f}_{;ab}\,+\,\Psi\,g_{ab}\,=(2\pi/\hbar \alpha)\,T_{ab} \label{Pso}
\ee
where $\Psi$ is an undetermined function. With the same argument given in section \ref{ees}, one then requires the conservation of the matter stress-energy tensor and use the contracted Bianchi identity to write the commutator of the covariant derivative as $2{v^c}_{;[ab]}=R_{abd}{}^c\,v^d$. In this way,
one finds
\be
%(\bold{f}\,R_{ab}\,-\,\bold{f}_{;ab})^{;a}\,=\left(\frac{1}{2}\, \mathcal{L}\,-\,\Box\,\bold{f} \right)_{,b},
(\bold{f}\,R_{ab}\,-\,\bold{f}_{;ab})^{;a}\,=\left(\frac{1}{2}\, f \,-\,\Box\,\bold{f} \right)_{,b}, \label{eq:int}
\ee
and thereby
\be
%\Psi\,=\left(\Box\,\bold{f}\,-\,\frac{1}{2}\, \mathcal{L} \right).\label{Psi}
\Psi\,=\left(\Box\,\bold{f}\,-\,\frac{1}{2}\, f  \right).\label{Psi}
\ee

Eventually, equation (\ref{Psi}), together with (\ref{Pso}) exactly leads, as expected, to the field equations of $f(R)$ gravity
\be
%\bold{f}\,R_{ab}\,-\,\bold{f}_{;ab}\,+\,\left(\Box\,\bold{f}\,-\,\frac{1}{2}\, \mathcal{L} \right)\,g_{ab}\,=\frac{2\pi}{\hbar \a}\,T_{ab}. \label{steos}
\bold{f}\,R_{ab}\,-\,\bold{f}_{;ab}\,+\,\left(\Box\,\bold{f}\,-\,\frac{1}{2}\, f \right)\,g_{ab}\,=\frac{2\pi}{\hbar \a}\,T_{ab}. \label{steos}
\ee
with the identification $\a=(4\hbar G)^{-1}$. In \cite{Eling:2006}, the same result was obtained starting from the entropy balance relation \eqref{ebal}, assuming $\sigma=0$, and then identifying the extra entropy term in $\theta$ in the second line of \eqref{eq:ds} with a suitable internal entropy term. There, it was also shown that the alternative route of keeping the $\theta^2$ in equation \eqref{eq:int} is not compatible with the conservation of the matter energy-momentum tensor.

Indeed, following the previous discussion, the above term is expected (together with a shear dependent term) as an unavoidable contribution related to the irreversible sector of the generalized Clausius relation \eqref{unhe} 
\be 
\d N=-\, \int_H \vS\,d\lambda\, \lambda \,(\alpha\,\bold{f})\,\left[\,\frac{3}{2}\,\theta^2\,+\,\norm\sigma\norm^2\right]_p, \label{kk}
\ee
which, as explained in section \ref{ne-th}, identifies the internal entropy production terms of the system.

As expected, the internal entropy in (\ref{kk}) now shows contribution both from scalar and tensorial degrees of freedom. Indeed, by using the same argument as in the GR case, we again have a natural interpretation for the expression in (\ref{kk}) as the dissipative function of the system. 

The shear squared contribution is equivalent to the one found for GR, with a shear viscosity coefficient which now takes a factor $\bold{f}$,
\be
\eta=\frac{\hbar\,\alpha\,\bold{f}}{4\pi}, \label{relofr}
\ee
as a consequence of the UV cut-off chosen for the area entropy relation.

On the other hand, the internal entropy contribution due to the scalar degree of freedom is now given by 
\be 
d_iS_{\theta}=-\, \int_H \vS\,d\lambda\, \lambda \,(\alpha\,\bold{f})\,\frac{3}{2}\,\theta_p^2. \label{scal}
\ee
By making use of the kinematical analogy described in section \ref{nccm}, and by expressing the above equation in the Killing frame, one is naturally led to define the bulk viscosity $\zeta$ as
\be
\frac{\zeta}{T}=\frac{3\,\a\, \bold{f}}{2\kappa} , \label{zeta}
\ee
that is $\zeta= 3\,\hbar\,\a\, \bold{f}/4\pi$, as already found in \cite{Eling:2006}.

\subsection{Gravitational dissipation in scalar-tensor gravity}

In order to give a physical interpretation to (\ref{scal}), one can use the equivalence between $f(R)$ and scalar-tensor gravity, thereby interpreting $\bold{f}$ as an effective massive dilaton.

The action for $f(R)$ gravity is given by
\be 
\mathcal{S}= \frac{\hbar \a}{4\pi}\int d^4x \sqrt{-g}\, f(R)\,+\, \mathcal{S}_{mat} \label{fr}.
\ee
By introducing an auxiliary field $\phi\equiv \mathbf{f}(R)$~\footnote{Note that the fact that in $f(R)$ gravity the associated scalar field is not a generic spacetime function but rather just of $R$, makes it possible to derive a closed system of equations without having to assume the equations of motion of the scalar field separately.}
 and assuming $f''(R)\ne 0$ for all $R$, one can take $V(\phi)$ as  the Legendre transform of $f(R)$ so that $R = V'(\phi)$, thereby rewriting the expression in (\ref{fr}) as
\be 
\mathcal{S}= \frac{\hbar \a}{4\pi}\int d^4x \sqrt{-g}\, \left[\phi\,R\,+\,V(\phi)\right]\,+ \mathcal{S}_{mat} .\label{act}
\ee
The Euler-Lagrange equations, in the Jordan frame, take the form
\begin{eqnarray} 
\phi \left( R_{ab} - \frac{1}{2}g_{ab} R \right)\, + \,\left(g_{ab}\Box -\nabla_a \nabla_b \right) \phi \,+ \\ \nn
+\, \frac{1}{2} g_{ab}V(\phi) =\frac{2\pi}{\hbar \a}\, T_{ab}, \label{el}
\end{eqnarray}
equivalent to field equations given in (\ref{steos}).

In this frame, by using the relation (\ref{rela}), one can express the dissipated energy coupled to the bulk and shear viscosity in (\ref{scal}), in terms of the auxiliary scalar field $\phi$
\be 
T\d N =- \int_H \vS\,d\lambda\,\lambda \,(\alpha\,\phi)\,T\,\left[ \frac{3}{2}\,\phi^{-2}\,\dot{\phi}^2+\norm\sigma\norm^2\right]_p. \label{ooh}
\ee

Similarly to the GR case, one expect that relation (\ref{ooh}) expresses some purely gravitational energy loss for the system, this time involving both scalar and tensorial fluxes. 
The interpretation of the term related to the shear is straightforward as it is clearly the generalization to a scalar-tensor theory of the tidal heating already obtained for the GR case.
More problematic is the interpretation of the bulk viscosity (purely scalar) contribution as no equivalent derivation as that for the tidal heating term in GR has be performed (to our knowledge) for scalar-tensor theories of gravity.

In this direction, as a first step, one can look for the effective field source terms which drive the local deformation of the null horizon congruence. Again, we consider the effective expansion $\tilde{\theta}$ as the suitable quantity to describe the scalar perturbations of the horizon given that $\tilde{\theta}=0$ is the condition for equilibrium.

Let us then use equation (\ref{etr}) as the effective expansion rate for the horizon bundle at $p$.  Starting from this equation, where now $\bold{f}$ is substituted by $\phi$, then using the Raychaudhuri equation and the Jordan frame field equation given in (\ref{el}), one gets
\begin{eqnarray}
\dot{\tilde{\theta}}_p\,=\,-\frac{\theta_p^2}{2}\,-\,\norm\sigma\norm_p^2\,-\,(2\pi/\hbar \a)\,\phi^{-1}\,T_{ab}\ell^a \ell^b\,-\\ \nn
-\,\phi^{-2}\,\nabla_a\,\phi\nabla_b\phi\, \ell^a \ell^b .\,\,\,\,\,\,\,\,\,\,\,\,\,\,\,\,\,\,\,\,\,\,\,\,\,\,\,\,\,\,\,\,\,\,\,\,\,\,\,\,\,\,\,\,\,\,\,\,\,
\end{eqnarray}
Now, since we are interested only the scalar contributions to $ \dot{\tilde{\theta}}_p$, we can set $T_{ab}=0$ and $\norm\sigma\norm_p^2=0$. Then, by using the equilibrium relation in the new frame $\theta=-\,\phi^{-1}\dot{\phi}$, we are left with
\be
\dot{\tilde{\theta}}_p\,=-\frac{3}{2}\,\phi^{-2}\,\nabla_a\,\phi\nabla_b\phi\, \ell^a \ell^b=-\,\frac{3}{2}\,\phi^{-2}\,\dot{\phi}^2.\label{til}
\ee
We see now that equation (\ref{til}) identifies the quantity $(3/2)\,\phi^{-2}\,\dot{\phi}^2$ as what one might define as the gravitational energy flux associated with the solely scalar field degrees of freedom.

In conclusion, we can now provide  a clean interpretation of the viscous terms in \eqref{ooh} as those representing the thermal system internal energy loss due to both scalar and tensor gravitational energy fluxes through the horizon. 
In particular, by moving to the Killing frame, (\ref{ooh}) can be rewritten as
\be 
T\d N=\frac{1}{8 \pi G}\int_H \vS\,dv \,\left[\frac{3}{2} \phi^{-1}\,\dot{\phi}^2+\phi \norm\sigma\norm^2\right]_p, \label{sfl}
\ee
where we have set $\a=(4\hbar G)^{-1}$ as required by the equations of motion \eqref{steos}. The striking similarity of the above expression with the energy loss rate due to the gravitational radiation in scalar-tensor gravity (see e.g.~equation (10.135) of \cite{Will}) further reinforces the above suggested interpretation.  Furthermore, we now see that also for the case of $f(R)$ gravity the relations \eqref{relofr}, \eqref{zeta} plus the above interpretation of \eqref{sfl} provide a correlation between the transmission coefficient of the gravitational energy through the horizon and the horizon congruence viscosity.

\section{Discussion}

In non-equilibrium spacetime thermodynamics the viscous dissipative effects appear to be naturally associated with purely gravitational energy fluxes. Their association to the irreversible/dissipative sector of the theory strongly suggests an interpretation of their nature as, non-local, internal heat flows associated with the internal spacetime degrees of freedom and clarifies why in GR a local, background independent, description of gravitational waves is precluded.

Noticeably, in order to recover the field equations one always need to effectively isolate these dissipative contributions by neatly separating the reversible and irreversible sectors of the constitutive equation. The analogy between the horizon null congruence and a continuous medium allows to recognize the natural terms related to the irreversible sector of the entropy balance. 
This effective separation of the reversible and irreversible regimes is further supported, at least in GR, by the fact that the energy contributions occurring in the equilibrium constitutive relations  have a local nature, being always related to the gravity field sources (Ricci curvature), whereas the non-equilibrium terms are intrinsically non-local and related to those curvature components which are independent from the sources distribution (Weyl curvature). This actually shows that the thermodynamical derivation of the gravitational  field equations is very general as it is sensitive to the whole spacetime curvature.\footnote{The situation in $f(R)$ gravity is less clear due to the presence of the scalar field dissipative contribution in  \eqref{sfl}. It has, however, to be noticed that this is not a part of the stress-energy tensor of the scalar field for the scalar-tensor theory equivalent to $f(R)$.}

%%%%%%%%%

However, a different issue is the interpretation of the internal entropy production terms related to dissipation with respect to a particular spacetime solution. While the association between internal entropy and allowed form of gravitational fluxes seems quite clear (e.g. we showed that the energy dissipated in GR {\em coincides} exactly the Hartle-Hawking tidal heating term), it might seem however puzzling that the arbitrariness in the choice of $B$ allows for non-zero shear and expansion of the null congruence (and hence for internal entropy production terms) even, for example, if one imagine to have performed the local Rindler wedge construction in a Minkowski spacetime. 

In fact, the thermodynamical approach is providing us just with the constitutive equations of the thermal system associated with local Rindler wedge, not of the spacetime in which the latter is constructed.  The arbitrariness of the choice of $B$ (and hence of the thermal system properties) implies that such equations will at most characterize the structure of the gravitational theory selected by the entropy-area relation (the EP formulation). In this sense they will not be associated with physical fluxes or curvatures of the spacetime as a whole.
Hence, the possible presence of internal entropy terms, even when the local Rindler wedge is constructed in flat spacetime, does not imply that the latter can be seen as a system in a non-equilibrium state.

Of course, one might take an alternative point of view, and claim that the above discussion actually shows an intrinsic limitation of the standard construction adopted here, as in~\cite{Jacobson:1995ab,Eling:2006}. In addressing this issue, a possibility could consist in a further characterization of the local Rindler wedge construction. In fact, one might choose to construct the 2-surface $B$ in such a way that it will be sensitive to the local curvature at $p$ and reduce to a plane in the flat spacetime case (i.e. $B$ would lead generically to a non-zero $\theta_p$ and $\sigma_p$ but would also reduce to the standard bifurcation surface at $p$ for a Rindler wedge, whose orthogonal null congruence has $\theta_p=\sigma_p=0$, in the flat spacetime limit). For example, this could be achieved by constructing $B$ as a totally geodesic 2-dimensional spacelike sub-manifold of the spacetime passing through $p$~\cite{Oneill}. (That is, any geodesic passing through $p$ and there tangent to $B$ would have to be completely contained in $B$.) Within this alternative approach, while all the formula would still pertain to the thermodynamical behavior of the local Rindler wedge at $p$~\footnote{E.g. one can talk about dissipation only with reference to the local Rindler wedge as spacetime as a whole as to be seen as a conservative system.}, they would now be able to specialize to a  specific spacetime choice and hence link the dynamical behavior of the wedge to the actual local matter-curvature content of the chosen spacetime. 

We do not see at the moment a decisive argument to go in one sense or the other. All in all, the whole point of the thermodynamical approach is not to provide an instrument able to reconstruct the kind of spacetime one is living in. Rather it is aimed to put in evidence the thermodynamical structure and the internal degrees of freedom of gravitational theories. In this sense the traditional construction, with an arbitrary $B$, seems sufficient. We plan however to further explore this issue in future work.

Another important aspect of this approach highlighted by this work has to do with the role that the different Equivalence Principle formulations play in selecting the possible gravitational dynamics.

The \emph{Strong} Equivalence Principle implies, for a generic choice of $B$, $\theta_p=0$, thereby leading to the equations of motion of GR (with irreversible level only corresponding to tensorial gravitational fluxes).\footnote{We suspect that the same formulation, together with the restriction of patches $B$ such that the null congruence shear is zero at any point $p$ (taken now as a defining property of the theory), may imply a restriction to the conformally invariant sector of GR (Einstein equations plus the condition of everywhere vanishing Weyl), which is basically equivalent to Nordstr\"om gravity.}

The \emph{Einstein} Equivalence Principle, does not fix either $\theta$ nor $\sigma$, thereby allowing for a generalized theory of gravity. For the simple case of $f(R)$, we showed how the scalar degree of freedom in fact produces a dissipative contribution which is actually associated with some purely gravitational scalar energy flux through the system boundary (and which is not part of the SET of the scalar field). 

The presence of an unavoidable purely scalar gravitational energy flux seems to indicate that GR and scalar-tensor are truly separated theories. However, the very same thermodynamical approach, might also suggest to look at such gravitational theories as different regimes of some more general effective description of gravity (in the same way as a incompressible flows can be seen as a special regime of a general compressible fluid).  This seems an intriguing possibility worth further investigation as it might lead to a unified framework which associates different gravitational theories with different hydrodynamical regimes of the analogue flow associated with the horizon null congruence.  

For what regards the spacetime viscosity, the quantities $\eta$ and $\zeta$ are found to be always related to the UV cut-off scale of the theory through the entropy density. The UV cut-off scale can be identified as the scale at which the gravitational action is of the order of the quantum of action $\hbar$, therefore where the horizon is subject to some quantum fluctuations, the so called horizon \emph{zitterbewegung}. This naturally suggest to interpret the spacetime dissipative effects as a consequence of an underlying fluctuation behaviour of spacetime at the UV cut-off scale, as already considered in seventies by Candelas and Sciama \cite{candela} and nowadays explored in different terms in the AdS/CFT correspondence context \cite{Eling:2008,poli}.  In this sense, it is interesting to note that while in principle  $\a$ can {\em a priori} depend on the nature of the quantum fields and their interactions, consistency of the gravitational equations derived from the thermodynamical approach, implies a trivial relation to the Newton constant. This might suggest an underlying microscopic interpretation of gravity along the ideas of induced gravity (see e.g.~\cite{Jacobson:1994} for a related discussion).

Finally, the separation between reversible and irreversible sectors of gravity seems to indicate that dissipation can occur just at the gravitational level, without any contribution from matter, however phenomena like the Hawking radiation, with its back-reaction on the geometry (black hole evaporation), seems to suggest that this separation might be a by product of the test field treatment so far adopted.

\section*{Acknowledgments}
The authors wish to thanks S.~Sonego for illuminating discussions and comments. C.~Eling, L.~Sindoni and T.~Sotriou for useful comments and T.~Jacobson for constructive criticism on a preliminary version of the manuscript.

\end{document}